\documentstyle[11pt,etendu]{article}

\begin{document}

\bibliographystyle{plain}    


\title{On the Automated Classification of Web Sites}

\author{John M. Pierre \\ ~ \\
   {\large Interwoven, Inc.} \\
   {\large 101 2nd Street, 4th Floor} \\
   {\large San Francisco, California USA} \\
   {\large jpierre@interwoven.com}}


\sernum{0}                   

\serznum{000}                

\trdate{February 4}           

\tryear{2001}                

\trvol{6}                    

\trauthor{John M. Pierre}    

\newcommand{\emp}[1]{{\em #1}}


\maketrtitle                 


\trtitlepage                 

\begin{trabstract}           

In this paper we discuss several issues related to automated
text classification of web sites.  We analyze the nature of web
content and metadata in relation to requirements for text
features.  We find that HTML metatags are a good source of text
features, but are not in wide use despite their role in search
engine rankings.
We present an approach for targeted spidering
including metadata extraction and opportunistic crawling of
specific semantic hyperlinks.  We describe a system for
automatically classifying web sites into industry categories
and present performance results based on different combinations
of text features and training data.  This system can serve as the
basis for a generalized framework for automated metadata creation.

\end{trabstract}

\trabstractpage              

\small                       


\section{Introduction}
\label{intro}

There are an estimated 1 billion pages accessible on the world wide web with
1.5 million pages being added daily.  
Describing and organizing this vast amount of content is essential
for realizing the web's full potential as an information resource.
Accomplishing this in a meaningful way will require consistent
use of metadata and other descriptive data structures such as
semantic linking\cite{bernerslee}.
Categorization is an important ingredient as is 
evident from the popularity of web
directories such as Yahoo!\cite{yahoo}, Looksmart\cite{looksmart}, and the 
Open Directory Project\cite{dmoz}.  However these resources have been 
created by large teams of human editors
and represent only one type of classification scheme that, while widely
useful, can never be suitable to all applications.  Classification is a
fundamental intellectual task, and we take it as an
axiom that it is important and indeed essential for
organizing and understanding web content.

Automated classification is needed for at least two important reasons.
The first is the sheer scale of resources available on the web and their
ever-changing nature.  It is simply not feasible to keep up with
the fast pace of growth and change on the web
through a manual classification effort
without expending immense time and effort.
The second reason is that classification itself is a subjective
activity.  Different classification schemes are needed for different
applications.  No single classification scheme is suitable for
all applications.  Therefore different types of classification schemes,
representing different facets of knowledge, may need to be applied
in an ongoing fashion as new applications demand them. 
Domain specific classification
schemes, which can be quickly applied to large amounts of content using
automated methods, hold great
promise for generating effective metadata.

Classification should be considered within the larger context of
subject-based metadata.  Specific fields in metadata records often
correspond to different classification schemes.
The effective use of rich metadata will be important for establishing
and leveraging the power of the semantic web.  If web content shifts
from primarily text-based to primarily multimedia oriented,
metadata will become even more important.  Structured metadata can
serve as a driver for many applications such as knowledge based
search and retrieval, reasoning engines, intelligent agents,
and multi-faceted organization of information.  However metadata
creation can be tedious and time consuming.  Automated methods, such
as the one described in this paper, can be useful for facilitating
metadata creation.

In this paper we discuss some practical issues for applying methods of
automated classification to web content.  Rather than take a
one size fits all approach we advocate the use of targeted specific
classification tasks, relevant to solving specific problems.
In section \ref{theweb} we discuss the nature of web content
and its implications for automated categorization.  
Extracting good features that can accurately discrimintate between
different categories is an important part of any text categorization system.
While it is possible and desirable to exploit metadata in the 
current web environment, we find that its use is far from widespread.
In section \ref{setup} we describe a specialized system for 
automatically classifying
web sites into industry categories.   This system
can serve as a generalized framework for efficient automated 
categorization of web content that includes targeted spidering,
domain specific classification, and a trainable general purpose
text categorization engine.
In section \ref{results} we present the results of our controlled
experiments.  We show how text features extracted from different parts of
web pages effect classification accuracy, and demonstrate that metatags
provide the best results.  We also compare the use of training data
obtained from a different domain versus training data drawn from the
target domain.  We find that training examples taken from the
content to be classified give better results, but using training data
from a different domain can suffice in cases where assembling new
data from scratch is not feasible.
Related work is discussed in section \ref{relatedwork}. 
In section \ref{conclusions} we state our conclusions and make 
suggestions for further research.

\section{Text Categorization of Web Content}
\label{theweb}

The current state of the web differs markedly from the vision of the
semantic web as outlined by Tim Berners-Lee\cite{bernerslee}.  
While web content is machine
readable for the most part\footnote{The trend toward multimedia assets
puts the future of this assumption in some doubt, but dealing with
the problem of non-text information is beyond the scope of this paper.}, 
it is far from machine understandable.
Furthermore the ability for computers to understand written human
language is still quite limited at this point in time.  Therefore,
in this work we have adopted a text categorization approach that
relies heavily on word-based indexing and statistical classification, 
rather than
sophisticated natural language processing and knowledge-based inferencing.
This approach is capable of giving very good results in a way that is
robust and makes few assumptions about the content to be analyzed. This
is an important consideration given the heterogenous nature of web content.

One the main challenges with classifying web pages is the 
wide variation in their content and quality.  
Most text categorization 
methods rely on the existence of good quality texts, especially for 
training\cite{lewis92}.
Unlike many of the well-known collections typically studied in
automated text classification experiments (i.e. TREC, Reuters-22578, OSHUMED),
in comparison the web lacks homogeneity and regularness.  
To make matters worse,
much of the existing web page content is based in images, 
plug-in applications, or other non-text media.  The usage of metadata
is inconsistent or non-existent.  In this section we survey
the landscape of web content, and its relation to the 
requirements of text categorization systems.

\subsection{Analysis of Web Content}

In an attempt to characterize the nature of the content to be
classified, we performed a rudimentary quantitative analysis.
Our results were obtained by analyzing a collection of 29,998
web domains obtained from a random dump of the database
of a well-known domain name registration company.  
Of course these results
reflect the biases of our small samples and don't necessarily generalize to
the web as a whole, however they should be reflective of the issues
at hand.  Since our classification method is text based, it is important
to know the amount and quality of the text based features that typically
appear in web sites.  Existing standards for web content tend to be
\textit{de facto} and loosely enforced if at all.  
One convention that holds for the vast majority of web sites is that
the top level entry point is an HTML web page, so we take this to be our
primary source of text features.
Besides the body text
which is generally free form in a typical HTML page, it is
common to include a title and possibly a set of keywords and description
metatags.   One of the more promising sources 
of text features should be found in web page metadata.

In Table \ref{metawords} we show the percentage of web sites with a certain
number of words for each type of metatag.
We analyzed a sample of 19195 domains with live web sites and counted
the number of words used in the content attribute of the
\texttt{<META name=``keywords''>} and \texttt{<META name=``description''>} 
tags as well as \texttt{<TITLE>} tags.  We also counted free text
found within the \texttt{<BODY>} tag, excluding all other HTML tags.

\begin{table*}[!hp]
\caption{Percentage of Web Pages with Words in HTML Tags}
\label{metawords}
\begin{center}
\begin{tabular}{crrrr}
\hline
Tag Type & 0 words & 1-10 words & 11-50 words & 51+ words \\
\hline
Title & 4\% & 89\% & 6\% & 1\% \\
Meta-Description & 68\% & 8\% & 21\% & 3\% \\
Meta-Keywords & 66\% & 5\% & 19\% & 10\% \\
Body Text & 17\% & 5\% & 21\% & 57\% \\
\hline
\end{tabular}
\end{center}
\end{table*}

The most obvious source of text is within the body of the web page.
We noticed that about 17\% of top level web pages had no usable body
text.  These cases include pages that only contain frame sets,
images, or plug-ins (our user agent followed redirects whenever
possible).  Almost a quarter of web pages contained 11-50 words,
and the majority of web pages contained over 50 words.

Though title tags are common the amount of text is relatively small with
89\% of the titles containing only 1-10 words.
Also, the titles often contain only names or terms such as 
``home page'', which are not particularly helpful for subject classification.

Metatags for keywords and descriptions are used by several major search
engines, where they play an important role in the ranking and
display of search results.  Despite this, only about a third of
web sites were found to contain these tags.
As it turns out, metatags can be useful when they exist
because they contain text specifically intended to aid in the
identification of a web site's subject areas\footnote{The possibilities 
for misuse/abuse of these tags to improve search engine rankings are well 
known; however, we found these practices to be not very widespread in our 
sample and of little consequence.}.  Most of the time these metatags
contained between 11 and 50 words, with a smaller percentage containing
more than 50 words (in contrast to the number of words in the body
text which tended to contain more than 50 words).

The lack of widespread use of metatags, despite the apparent incentive
to improve search engine rankings, is instructive.  Since metadata 
is usually not part of the presentation of the content and its
benefit is somewhat intangible, it tends to be neglected.  Creating metadata
can be a tedious and unwelcome task.  Therefore methods to facilitate the
creation of quality metadata, especially automated methods, are greatly
needed.

\subsection{Good Text Features}
\label{goodfeatures}

Feature selection is an important part of building an automated
classification system.  Without a proper set of features, the
classifier will not be able to accurately 
discriminate between different categories.
The feature set must be sufficiently broad to acommodate the wide
variations that can occur even within instances of the same class.  On the
other hand the number of features needs to be constrained to reduce noise 
and to limit the burden on system resources.

In reference\cite{lewis92} it is argued that for the purposes of automated
text categorization, features should be:
\begin{enumerate}
\item Relatively few in number
\item Moderate in frequency of assignment
\item Low in redundancy
\item Low in noise
\item Related in semantic scope to the classes to be assigned
\item Relatively unambiguous in meaning
\end{enumerate}

Due to the wide variety of purpose and scope of current web content,
items 4 and 5 are difficult requirements to meet for most
classification tasks.  For subject
classification, metatags seem to meet those requirements better
than other sources of text such as titles and body text.  However
the lack of widespread use of metatags is a problem if 
coverage of the majority of web content is desired.  In the long term, 
automated categorization could really benefit if greater
attention is paid to the creation and usage of rich metadata and
explicit semantic structures,
especially if the above requirements are taken into consideration.
In the short term, one must implement a strategy for obtaining
good text features from the existing HTML and natural language
cues that takes the above requirements as well as the goals
of the classification task into consideration.  Techniques for shallow
parsing and information extraction are useful in this regard.

\section{Experimental Setup}
\label{setup}

We constructed a full scale automated classification system and
performed several experiments using real world data in order to
gauge system performance and test ideas.
The goal of our targeted domain specific task was to 
rapidly classify web sites (domain names)
into broad industry categories. In this section we describe the
main ingredients of our classification experiments including the data,
architecture, and evaluation measures.

\subsection{Classification Scheme}

The categorization scheme used was the
top level of the 1997 North American Industrial Classification System
(NAICS) \cite{naics}, which consists of 21 broad industry categories
shown in Table \ref{tnaics}.

\begin{table*}[!htbp]
\caption{Top level NAICS Categories}
\label{tnaics}
\begin{center}
\begin{tabular}{cl}
\hline
NAICS code & NAICS Description \\
\hline
11 & Agriculture, Forestry, Fishing, and Hunting \\
21 & Mining \\
22 & Utilities \\
23 & Construction \\
31-33 & Manufacturing \\
42 &  Wholesale Trade \\
44-45 &  Retail Trade \\
48-49 &  Transportation and Warehousing \\
51 &  Information \\
52 &  Finance and Insurance \\
53 &  Real Estate and Rental and Leasing \\
54 &  Professional, Scientific and Technical Services \\
55 & Management of Companies and Enterprises \\
56 &  Administrative and Support, \\
   &  Waste Management and Remediation Services \\
61 & Educational Services \\
62 & Health Care and Social Assistance \\
71 & Arts, Entertainment and Recreation \\
72 & Accommodation and Food Services \\
81 & Other Services (except Public Administration) \\
92 & Public Administration \\
99 & Unclassified Establishments \\
\hline
\end{tabular}
\end{center}
\end{table*}

Some of our resources had been previously classified using the older
1987 Standard Industrial Classification (SIC) system.  In these cases
we used the published mappings\cite{naics} to convert all
assigned SIC categories to their NAICS equivalents.  The full
NAICS has six levels of hierarchy and contains
several thousand subcategories.  For our experiments all lower level
NAICS subcategories were generalized up to the appropriate
top level category (though the entire classification scheme could
have been utilized by our system if a finer grained categorization
was desired).

NAICS and SIC are examples of authoritative controlled vocabularies.  
Using a published standardized classification scheme can be a good idea 
in order to take advantage of the many person hours of time it takes
to construct something like this.  In addition, it may be possible to
take advantage of existing content already classified by the scheme as
a source of training data.

\subsection{Targeted Spidering}
\label{spider}

Based on the results of section \ref{theweb}, it is obvious that selection
of adequate text features is an important issue and certainly
not to be taken for granted.  To balance the
needs of our text-based classifier against the speed and storage limitations of
a large-scale crawling effort, we took an approach for spidering
web sites and gathering text that was targeted to the classification task
at hand.  

In some preliminary tests we found the best classifier accuracy
was obtained by using only the contents of the keywords and
description metatags as the source of text features.  Adding
body text decreased classification accuracy.  However, due to
the lack of widespread usage of metatags limiting ourselves
to these features was not practical, and other sources of
text such as titles and body text were needed to provide
adequate coverage of web sites.  Therefore our targeted spidering approach
attempted to gather the higher quality text features from metatags
and only resorted to lower quality texts if needed.

Our opportunistic spider began at the top level page of the web site
and attempted to extract useful text from metatags and titles
if they exist, and then followed links for frame sets if they existed.  
It also followed any hyperlinks
that contained key substrings in their anchor text
such as \emph{product}, \emph{services},
\emph{about}, \emph{info}, \emph{press}, and \emph{news}, and again
looked for metatag content in those pages.  
These substrings were chosen based on
an \emph{ad hoc} frequency analysis and the assumption that they tend to
point to content that is useful for deducing an industry classification.
Only if no metatag content was found did the spider
gather the actual body text of the web page.  All extracted text was
concatenated into a single representative document for the site
that was submited to the classification engine.
For efficiency we ran several spiders in parallel, each working
on different lists of individual domain names.

What we were attempting to do by following a restricted set of hyperlinks,
was to take advantage of the
current web's \emph{implicit} semantic structure.
One the advantages of moving towards an \emph{explicit} semantic
structure for hypertext documents\cite{bernerslee} is that an
opportunistic spidering
approach could really benefit from a formalized description of the
semantic relationships between linked web pages.  This would allow
spiders to more easily find the most relevant resources without having to
crawl the entire network of the web.

\subsection{Test Data}

From our initial list of 29,998 domain names we used our targeted spider
to determine which sites were live and extracted
text features using the approach outlined in section \ref{spider}.
Of those, 13,557 domain names had usable text content and were pre-classified
according to one or more industry categories\footnote{Industry classifications
for domain names were provided by InfoUSA and Dunn \& Bradstreet.}.  From 
this set of data we drew samples for training, testing and validation.

\subsection{Training Data}
\label{ts}

We took two approaches to constructing training sets for our
classifiers.  In the
first approach we used a combination of 426 NAICS category labels 
(including subcategories) and 1504 U.S. Securities and Exchange Commission
(SEC) 10-K filings\footnote{SEC 10-K filings are annual reports
required of all U.S. public companies that describe business
activities for the year.  Each public company is also
assigned an SIC category.} 
for public companies\cite{dolin99} as training examples.  
In the second approach we used a set of 3618 pre-classified
domain names along with text for each domain obtained using our spider.

The first approach can be considered as using ``prior knowledge''
obtained in a different domain.  It is interesting to see how knowledge from
a different domain generalizes to the problem of classifying web sites.  
Furthermore it is
often the case that training examples can be difficult to obtain (thus
the need for an automated solution in the first place).  The
second approach is the more conventional classification by example.
In our case it was made possible by the fact that our database
of domain names was pre-classified according one or more industry categories.

\subsection{Classifier Architecture}

Our text classifier consisted of three modules: the targeted spider for 
extracting text features associated with a web site, 
an information retrieval engine for comparing queries to
training examples, and a decision algorithm for assigning categories.

Our spider was designed to quickly process a large database of
top level web domain names (e.g. domain.com, domain.net, etc.).
As described in section \ref{spider} we implemented an opportunistic
spider targeted to finding high quality text from pages that described
the business area, products, or services of a commercial web site.  
After accumulating text features, a query was submitted to the
text classifier.  The domain name and any automatically
assigned categories were logged in a central database.
Several spiders could be run in parallel for efficient use of system
resources.

Our information retrieval engine was based on Latent Sematic Indexing 
(LSI)\cite{lsi}.  LSI is a variation of the vector space model of
information retrieval that uses the technique of singular value
decomposition (SVD) to reduce the dimensionality of the vector space.
Words that tend to co-occur in the same document share large projections
along directions in the reduced space.  Theoretically this reduces
noise due to redundant or spurious word usage, and automatically 
derives relationships
between words and the inherent concepts.  Cosine similarity is computed
in the reduced vector space, which amounts to concept based matching rather
than word based.  For example queries containing the word ``car'' will
match documents containing only the word ``automobile'' provided the
relationship between the words and concept has been established in the corpus.

In a previous work\cite{dolin99} it was shown that LSI provided better
accuracy with fewer training set documents per category than standard
TF-IDF weighting.  Queries were compared to training
set documents based on their cosine similarity, and a ranked list of
matching documents and scores was forwarded to the decision module.

In the decision module,
we used a K-nearest neighbor algorithm for ranking categories and
assigned the top ranking category to the web site.  This type of classifier
tends to perform well compared to other methods\cite{yang}, is robust,
and tolerant of noisy data (all are important qualities when dealing with
web content).  In addition the algorithm is capable of producing good
results even when the amount of training data is limited.  The decision
module also is responsible for thresholding and presenting the final
set of automatically assigned categories.

\subsection{Evaluation Measures}

System evaluation was carried out using the standard precision,
recall, and F1 measures\cite{rijsbergen}\cite{lewis91}.  
Precision is the number of correct categories assigned divided
by the total number of categories assigned, and serves as a measure
of classification accuracy.  The higher the precision the smaller the amount
of false positives.  Recall is the number of correct categories
assigned divided by the total number of known correct categories.
Higher recall means a smaller amount of missed categories.  In theory,
scores of 1 are desirable for both precision and recall.  In practice
even human assigned classifications may only achieve scores between
0.7 and 0.9, depending on the classification task.  This is because
to some extent classification is a subjective task and there are
usually ``grey areas'' in a classification scheme.

The F1 measure combines precision and recall with equal importance
into a single parameter for optimization and is defined as
\begin{equation}
F1 = \frac{2 P R}{P + R}
\end{equation}
where P is precision and R is recall.

We computed global estimates
of system performance using both micro-averaging (results are computed
based on global sums over all decisions) and 
macro-averaging (results are computed on a per-category basis,
then averaged over categories).  Micro-averaged
scores tend to be dominated by the most commonly used categories,
while macro-averaged scores tend to be dominated by the performance
in rarely used categories.  This distinction was relevant to our problem,
because it turned out that the vast majority of commercial web sites
are associated with the Manufacturing (31-33) category.

\section{Results}
\label{results}

In our first experiment we varied the sources of text features
for 1125 pre-classified web domains.  We
constructed separate test sets
based on text extracted from the body text, metatags 
(keywords and descriptions),
and a combination of both.  The training set consisted of SEC documents
and NAICS category descriptions.
Results are shown in Table \ref{ptf}.

\begin{table}[!htbp]
\caption{Performance vs. Text Features}
\label{ptf}
\begin{center}
\begin{tabular}{cccc}
\hline
Sources of Text & micro P & micro R & micro F1 \\
\hline
Body & 0.47 & 0.34 & 0.39 \\
Body + Metatags & 0.55 & 0.34 & 0.42 \\
Metatags & 0.64 & 0.39 & 0.48 \\ 
\hline
\end{tabular}
\end{center}
\end{table}

Using metatags as the only source of text features resulted in
the most accurate classifications.  Precision decreases noticeably
when only the body text was used.  It is interesting that including
the body text along with the metatags also resulted in less accurate
classifications.  These results influenced the design of our spider
which extracted metatags first and foremost, while only grabbing
body text as a last resort.
The usefulness of metadata as a source of high quality
text features should not be suprising since it meets most of the
criteria listed in \ref{goodfeatures}.

In our second experiment we compared classifiers constructed from
the two different training sets described in section \ref{ts}.  
The results are shown in Table \ref{pts}.

\begin{table*}[!htbp]
\caption{Performance vs. Training Set}
\label{pts}
\begin{center}
\begin{tabular}{ccccccc}
\hline
Classifier & micro P & micro R & micro F1 & macro P & macro R & macro F1\\
\hline
SEC-NAICS & 0.66 & 0.35 & 0.45 & 0.23 & 0.18 & 0.09 \\
Web Pages & 0.71 & 0.75 & 0.73 & 0.70 & 0.37 & 0.40 \\
\hline
\end{tabular}
\end{center}
\end{table*}

The SEC-NAICS training set achieved respectable micro-averaged scores,
but the macro-averaged scores were low.  One reason for this is that
this classifier generalizes well in categories that are
common to the business and web domains (31-33, 23, 51),
but has trouble with recall in
categories that are not well represented in the business domain
(71, 92) and poor precision in categories that are not as common in the web
domain (54, 52, 56).

The training set constructed from web site text performed better
overall.  Macro-averaged recall was much lower than micro-averaged
recall.  This can be partially explained by the following example.
The categories Wholesale Trade (42) and Retail Trade (44-45) have
a subtle difference especially when it comes to web page text
which tends to focus on products and services delivered rather
than the Retail vs. Wholesale distinction.  In our training set, category
42 was much more common than 44-45, and the former tended to be assigned
in place of the latter, resulting in low recall for 44-45.  Other
rare categories also tended to have low recall (e.g. 23, 56, 81).

\section{Related Work}
\label{relatedwork}

Some automatically-constructed, large-scale web directories have
been deployed as commercial services such as 
Northern Light\cite{northernlight},
Inktomi Directory Engine\cite{inktomi}, Thunderstone
Web Site Catalog\cite{thunderstone}.  Details about these
systems are generally unavailable because of their proprietary
nature.  It is interesting that these directories tend not to
be as popular as their manually constructed counterparts.

A system for automated discovery and classification of domain
specific web resources is described as part of the DESIRE II 
project\cite{desire1}\cite{desire2}.  Their classification
algorithm weights terms from metatags higher than titles and
headings, which are weighted higher than plain body text.
They also describe the use of classification software as a
topic filter for harvesting a subject specific web index.
Another system, Pharos (part of the Alexandria Digital
Library Project), is a scalable
architecture for searching heterogeneous information sources
that leverages the use of metadata\cite{dolin96} and 
automated classification\cite{dolin98}.

The hyperlink structure of the web can be exploited for automated
classification by using the anchor text and other context
from linking documents as a source of text features\cite{attardi}.
Approaches to efficient web spidering\cite{cho}\cite{rennie} have
been investigated and are especially important for very large-scale
crawling efforts.

A complete system for automatically building searchable databases of
domain specific web resources using a combination of
techniques such as automated classification, targeted spidering, and
information extraction is described in reference\cite{mccallum}.

\section{Conclusions}
\label{conclusions}

Automated methods of knowledge discovery, including classification,
will be important for establishing the semantic web.
Classification is a basic intellectual task and is challenging to
automate due to its somewhat subjective nature.  However it is possible
to achieve results with automated methods that meet or exceed manual results.

A single classification 
scheme can never be adequate for all applications.
We advocate a pragmatic approach including targeted techniques and
specialized domain knowledge to
be applied to specific classification tasks.  The result is an
efficient and optimized system for the task at hand.
In this paper we described a practical system for automatically
classifying web sites into industry
categories that gives good results.  This type of system can
be applied to any domain specific classification scheme.  All that is needed
is to define the categories, assemble the training data,
and configure the spider to extract the appropriate features.  The spider
may be constructed to follow specific types of links, or extract sections
of web page content that are most useful for a given domain.

From the results in Table \ref{ptf} 
we concluded that metatags were the best source
of quality text features, at least compared to the body text.  However
by limiting ourselves to metatags we would not be able to classify the
large majority web sites.  Therefore we opted for a targeted spider
that extracted metatag text first, looked for pages that
described business activities, and
then degraded to other text only
if necessary.  It seems clear that text contained in structured 
metadata fields results in better automated categorization.  If the
web moves toward a more formal semantic structure as outlined by
Tim Berners-Lee\cite{bernerslee}, then automated methods can benefit.  
If more and different kinds of automated 
classification tasks can be accomplished more accurately, the
web can be made to be more useful as well as more usable.

Rich metadata for web content is a key to better searching,
better organization and managment of content, and improved
intelligent agents capable of discovering and acting
upon the knowledge embedded in the vast online resources.
However, as we have shown,
 creation of metadata remains a bottleneck despite strong
incentives such as better rankings in search engine results.  It
seems that the only way to ensure widespread use of quality metadata
is to make the process of metadata creation as painless as possible.
Automated methods that can reliably and accurately generate metadata
from existing content hold much promise in this regard.  Furthermore
metadata needs to be multi-faceted, current, and extensible.  Only
automated systems can keep pace with the rate of generation of new
web content that we see today.

We outline our basic approach
for building a targeted automated categorization solution for web
content:
\begin{itemize}
\item \textbf{Knowledge Gathering} - It is important to have a 
clear understanding
of the domain to be classified and the quality of the content involved.  The
web is a heterogenous environment, but within given domains patterns and
commonalities can emerge.  Taking advantage of specialized knowledge can
improve classification results.
\item \textbf{Targeted Spidering} - For each classification task 
different features
will be important.  However, due to the lack of homogeneity in web content,
the existence of key features can be quite inconsistent.  A targeted spidering
approach tries to gather as many key features as possible with as
little effort as possible.  In the future this type of approach can
benefit greatly from a web structure that encourages the use of 
metadata and semantically-typed links.  It would be interesting to
do a more detailed analysis of semantic spidering and its effect on
system performance.
\item \textbf{Training} - The best training data comes from the
domain to be classified, since that gives the best chance
for identifying the key features.  In cases where it's
not feasible to assemble enough training data in the target domain,
it may be possible to achieve acceptable results using training data
gathered from a different domain.  This can be true for web content
which can be unstructured, uncontrolled, immense, and hence difficult
to assemble quality training data.  However, controlled
collections of pre-classfied electronic documents can be obtained
in many important domains (financal, legal, medical, etc.) and
applied to automated categorization of web content.
\item \textbf{Classification} - In addition to being
as accurate as possible, the classification method needs to
be efficient, scalable, robust, and tolerant of noisy data.  Classification 
algorithms that utilize the link structure of the web, including
formalized semantic linking structures should be further investigated. 
\end{itemize}

Non-text content such as images, applets, plugins, music and video
are becoming more and more prevalent on the web.  Devising automated
methods that can deal with this kind of content is an important area
for further investigation.  Again, effective use of metadata can be a
good way to help manage these types of non-text assets.

Better acceptance of metadata is one key to the future of the semantic web.
However, creation of quality metadata is tedious and is itself a
prime candidate for automated methods.  A preliminary method such
as the one outlined in the paper can serve as the basis for
bootstrapping\cite{boot} a more sophisticated classifier that takes full
advantage of the semantic web, and so on.

\section{Acknowledgements}
\label{acknowledgements}
I would like to thank for Bill Wohler for
collaboration on system design and software implementation, and
Roger Avedon, Mark Butler, and Ron Daniel for useful discussions.  
Special thanks to Network Solutions Inc. for providing classified domain names.


\end{document}